\documentclass{article}
\usepackage{amsmath, graphicx}
\usepackage{cite}

\newcommand{\bfr}{\begin{flushright}}
\newcommand{\efr}{\end{flushright}}

\newcommand\ignore[1]{}
\newcommand\be{\begin{equation}}
\newcommand\ee{\end{equation}}
\newcommand\bea{\begin{eqnarray}}
\newcommand\eea{\end{eqnarray}}
\newcommand{\bdm}{\begin{displaymath}}
\newcommand{\edm}{\end{displaymath}}
\newcommand\nn{ \nonumber\\}

\def\be{\begin{equation}}
\def\ee{\end{equation}}
\def\bea{\begin{eqnarray}}
\def\eea{\end{eqnarray}}

    \usepackage[margin = 0.9 in]{geometry}
\begin{document}
\title{Conformal
Pomeron and Odderon in Strong Coupling\thanks{Presented by Chung-I Tan at  the Low x workshop, May 30 - June 4 2013, Rehovot and Eilat, Israel.}
}

\author{Richard  C. Brower
\\
{\small  Boston University, Boston MA 02215, USA}
\smallskip\\
Miguel Costa,  Marko Djuric
\\
{\small  Universidade do Porto, 4169-007  Porto, Portugal}
\smallskip\\
Timothy Raben and  Chung-I Tan
\\
{\small  Brown University,
Providence, RI 02912, USA}
\smallskip\\
}
\date{September 23, 2013}

\maketitle
\begin{abstract}
We discuss how exact conformal invariance in the strong coupling leads naturally through AdS/CFT correspondence to a systematic expansion for the Pomeron and Odderon intercepts  in power of $\lambda^{-1/2}$, with $\lambda=g^2N_c$ the 't Hooft coupling. We also point out  the  importance of confinement  for a realistic treatment of  DIS in the HERA energy range.  
\\
~
\\
\end{abstract}

\section{Introduction}
In the past decade overwhelming evidence has emerged for a conjectured duality between a wide class of gauge theories in d-dimensions and string theories on asymptotically $AdS_{d+1}$ spaces. 
It has been shown, in a holographic or AdS/CFT  dual description for QCD at high energies, the Pomeron can be identified with a reggeized Graviton in $AdS_5$~\cite{Brower2007, Brower2009a} and, similarly, an Odderon with a reggeized anti-symmetric Kalb-Ramond $B$-field~\cite{Brower2009}. This approach has been successfully applied to the study of HERA data~\cite{Aaron2010a}, both for  DIS at small-$x$~\cite{Brower2010} and for  deeply virtual Compton scattering (DVCS)~\cite{Costa:2012fw}.  More recently, this treatment has also been applied to the study of diffractive production of Higgs at LHC~\cite{Brower2012} as well as other near forward scattering processes.

In this talk, we first briefly describe ``Pomeron-Graviton" duality and  its application for deep inelastic scattering (DIS) at small-x and deep virtual Compton scattering (DVCS) to HERA data. We next turn to a discussion on Pomeron and Odderon intercepts in the conformal limit and their  relation to the anomalous dimensions.

\section{Pomeron-Graviton Duality and Applications:}\label{sec:pomeron}

 It can be shown for a wide range of scattering processes that the amplitude in the Regge limit, $s\gg t$, is dominated by Pomeron exchange, together with the associated s-channel screening correction, e.g., via eikonalization. We will use here a formulation based on gauge/gravity duality, or the $AdS/CFT$ correspondence, of which one particular example is the duality between $\mathcal{N} = 4$ SYM and Type-IIB string theory on $AdS_5 \times S^5$. This approach has the advantages of allowing us to study the strong coupling region, providing a unified soft and hard diffractive mechanism, and as we will see it also fits well the experimental data. 

Traditionally the Pomeron  has been modeled at weak coupling using perturbative QCD;  
in lowest order, a bare Pomeron was first identified by Low and Nussinov as a two gluon exchange corresponding to a Regge cut in the $J$-plane at $j_0 = 1$.   Going beyond the leading order, Balitsky, Fadin, Kuraev and Lipatov (BFKL) summed generalized two gluon exchange diagrams to first order in $\lambda = g^2 N_c$ and {\em all} orders in $(\lambda \log s)^n$, giving rise to the so-called BFKL Pomeron~\footnote{ See~\cite{Kotikov:2004er} and references cited therein.}, which corresponds to a $J$-plane cut  at $j_0 = 1+ \log (2) \lambda /\pi^2$.   

In a holographic approach, the weak coupling Pomeron is replaced by the ``Regge graviton'' in AdS space, as formulated by Brower, Polchinski, Strassler and Tan (BPST)~\cite{Brower2007} which has both hard components due to near conformality in the UV and soft Regge behavior in the IR.  Strong coupling corrections lower the intercept from $j=2$ to
\be
j_0 = 2 - 2 /\sqrt{\lambda}   \; .
\label{eq:BPST-intercept}
\ee
\begin{figure}
\begin{center}
\includegraphics[height=0.25 \textwidth,width=.35\textwidth]{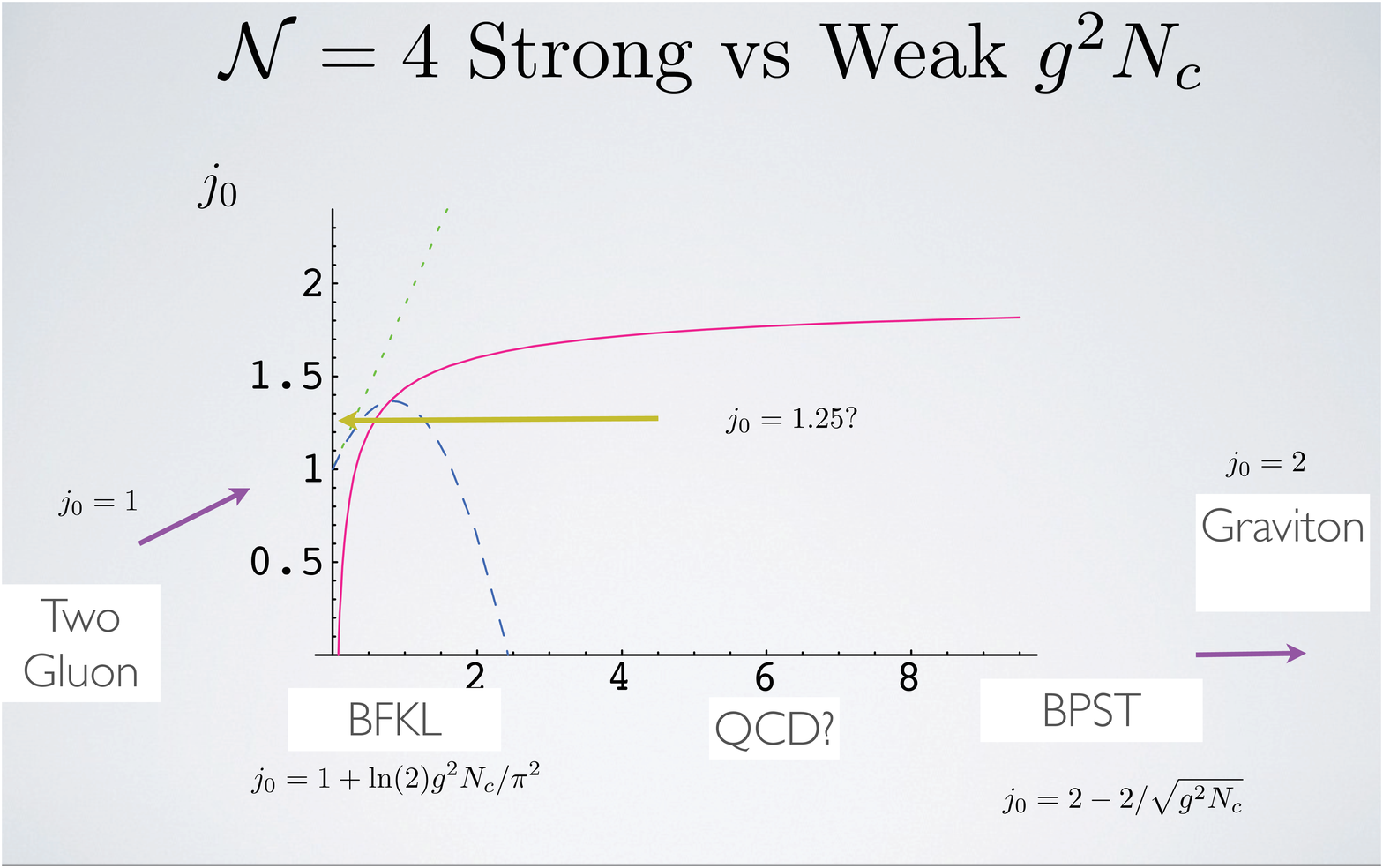}
\hskip 1.5cm 
\includegraphics[height=0.25 \textwidth,width=.35\textwidth]{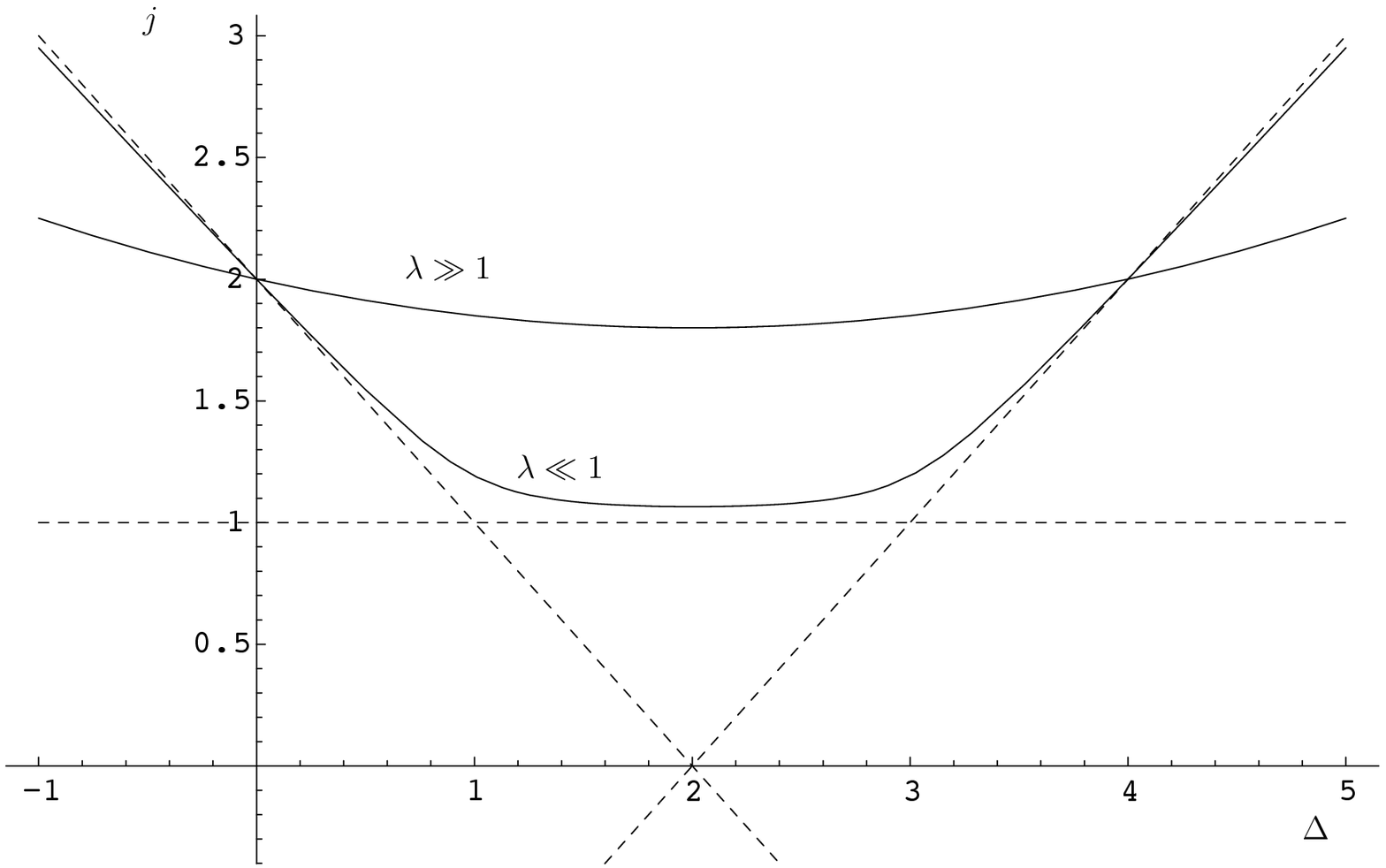}
\end{center}
\caption{ On the left,  (a), intercept  as a function of $\lambda$ for the BPST Pomeron (solid red) and for BFKL (dotted and dashed to first and second order in $\lambda$ respectively). On the right, (b), the conformal invariant $\Delta-j$ curve which controls both anomalous dimensions and the Pomeorn intercept.}
\label{fig:effective}
\end{figure}
In Fig.~\ref{fig:effective}a,  we compare the BPST Pomeron intercept   with the weak coupling BFKL intercept  for ${\cal N}=4$ YM as a function of 't Hooft coupling $\lambda$.  A typical phenomenological estimate for this parameter  for
QCD is about $j_0 \simeq 1.25$,  which suggests that the physics of diffractive scattering is in the cross over region between
strong and weak coupling.  A corresponding treatment for Odderons has also been carried out~\cite{Brower2009}. 
 
\subsection{BPST Pomeron Intercept in Conformal Limit:}
Let us begin by first examining briefly the concept of  a BPST Pomerorn in the general context of conformal  field theories (CFT).  A  CFT 4-point correlation function ${\cal A}=\langle \phi(x_1) \phi(x_2)  \phi(x_3)  \phi(x_4)\rangle$ can be analyzed in an operator product expansion (OPE) by summing over allowed primary operators ${\cal O}_{k,j}$, with integral spin $j$ and dimensions  $\Delta_k(j)$,  and their descendants.  It can be shown  that  the leading behavior in the Regge limit for the crossing even CFT correlation functions, appropriate for Pomeron exchange,  are controlled by a set of dominant single-trace primary operators ${\cal O}_{G,j}$, one for each $j$ and $j$ even, with conformal dimensions $\Delta_G(j)$. (We will return to this treatment in Sec.  \ref{sec:CFT}.) It is useful to express these scaling dimensions as 
\be
\Delta_G(j)=\tau_G+ j+\gamma_G(j),
\label{eq:GAD}
\ee
 where $\tau_G$ is the twist~\footnote{$\tau_G=2$ is the minimal twist which dominates the amplitude.} and $\gamma_G(j)$ are the anomalous dimensions. The lowest $j$ in this set has $j=2$, which is the energy-momentum tensor. Due to energy-momentum conservation, $\gamma_G(2)$ vanishes and $\Delta_G(4)=4$.   For simplicity, we will drop the subscript $G$ in what follows.   
 
 Using AdS/CFT in strong coupling, it was shown in \cite{Brower2007} that $\gamma(j)$ is analytic in $j$, so that one can expand  $
 \Delta(j) $ about $j=2$ as $  \Delta(j)   = 4 + \alpha_1(\lambda) (j-2)  + O((j-2)^2 )$, with the coefficient $\alpha_1(\lambda)= \sqrt\lambda/4 +  O(1)$  in the strong coupling limit.  Equivalently, one has an expansion  
\be
  (\Delta(j)-2)^2 = 4 + 4\alpha_1(\lambda) S + O(S^2)
    \label{eq:BPST-Basso}
\ee
where we have simplified the expression by introducing $S$ for  $j-2$. This notation is also useful for  the discussion in Sec.  \ref{sec:CFT} where we generalize the treatment to higher order in $1/\sqrt \lambda$ and also to the case of Odderons.
 
It was  stressed  in \cite{Brower2007} that  the $\Delta-j$ curve must be symmetric about $\Delta=2$ due to conformal invariance, and, by inverting $\Delta(j)$, one has
\be
 j(\Delta) = j(2) +\alpha_1(\lambda)^{-1}   (\Delta-2)^2 + O((\Delta-2)^4)
 \label{eq:j-Delta}
\ee
At large $\lambda$, the curve $j(\Delta)$   takes on a minimum at $\Delta=2$, as exhibited in  Fig.~\ref{fig:effective}b.   The Pomeron intercept is simply the minimum of $j(\Delta)$ curve at $\Delta=2$, that is, $j_{0}= j(2)$. In particular, it admits an expansion in $1/\sqrt \lambda$,
$
\alpha_P=j_{0}=2 -\frac{2 }{\lambda^{1/2}}+ \frac{a_{2} }{\lambda} +\frac{a_{3} }{ \lambda^{3/2}}+\frac{a_{4} }{\lambda^2}+\cdots.
$
The leading term corresponds to a graviton exchange and the first order correction comes from the classical contribution from string modes.  We will return to  higher order terms  in Sec. \ref{sec:CFT}.

 \ignore{
  $$
\alpha_O= j^{-}_0= 1 +\frac{a^-_{1}}{\lambda^{1/2}}+  \frac{a^-_2}{\lambda} +\frac{a^-_3}{ \lambda^{3/2}}+\frac{a^-_4}{\lambda^2}+\cdots.
$$
In \cite{}, two odderon solutions were found. Solution-A has $a^{-}_1= -8$.  The second solution, solution-B, surprisingly remains at 1 in the large $\lambda$ limit, i.e., has $a^{-}_1=0$. We have recently  extended the analysis of \cite{Brower:2008cy} to higher order, leading to  a similar expansion as Eq. (\ref{eq:P-intercept}), with  $a^{-}_2=-4, a^{-}_3=13$ for solution-A. Interestingly, for solution-B,  we find that 
$$
\alpha_O=1
$$
to all orders in $1/\sqrt\lambda$, consistent with that reported from a weak-coupling analysis.  More detailed analysis will be discussed  in a separate report.\footnote{BCDRT, in preparation.}
}

\subsection{Holographic Treatment of DIS and DVCS:}\label{sec:DIS}

In the holographic approach, the impact parameter space $(b_\perp, z)$ is 3-dimensional, where  $z \ge 0$ is the warped radial 5th dimension.  Conformal dilatations, ($z \rightarrow c z $ with $c$ a constant), take one from the UV boundary at $z = 0$ deep into the IR  $z = \mbox{large}$.    The near forward elastic amplitude  $A(s,t)$, where  $t=-q^2_{\perp}$, in  a transverse $AdS_3$ representation, 
$A(s,t)= \int d^{2}b \; e^{i \vec q \cdot \vec b} \int dzdz' {\widetilde A}(s,b,z,z')$, can be written in an eikonal form
\bea
\widetilde A(s,b,z,z')&=&2i s P_{13}(z)P_{24}(z')\big\{ 1- e^{i \chi(s,b,z,z')}\big\}  \; .\label{eq:A}
\eea
When expanded to first order in the eikonal function,  it leads to the contribution from exchanging a single Pomeron,  with 
$
\chi( s,b,z,z')=\frac{g_{0}^{2}}{2{s}}(\frac{R^{2}}{zz'})^2\mathcal{K}({s},b,z,z') 
$,
where  $\mathcal{K}({s},b,z,z') $, is the  BPST Pomeron kernel  in a transverse $AdS_3$ representation~\cite{Brower2007, Brower2009a}.   In the conformal limit, a simple expression for $\mathcal{K}({s},b,z,z') $ can be found~\cite{Brower2009a}. Confinement can next be introduced, e.g., via a hardwall model $z < z_{cut-off}$. The effect of saturation can next be included  via the full transverse $AdS_3$ eikonal representation (\ref{eq:A}).

An  important unifying feature for our treatment  is factorization in the AdS space.  For hadron-hadron scattering, $P_{ij}(z)= \sqrt{-g(z)} (z/R)^2 \phi_i(z) \phi_j(z) $  involves a product of two external normalizable wave functions for the projectile and the target respectively.  For scattering involving external currents, we can simply replace $P_{13}$ by product of the  appropriate unnormalized wave-functions.  

We  next make use of the fact that the DIS cross section can be related to the imaginary part of the forward amplitude via the optical theorem, $\sigma=s^{-1}{\rm Im} A(s,t=0)$.  Therefore, the AdS/CFT amplitude gives expressions for all structure functions $F_i$.   We shall focus here on  $F_2(x,Q^2) =\frac{Q^2}{\pi^2\alpha_{em}}(\sigma_T(\gamma^*p)+\sigma_L(\gamma^*p))$. In the conformal limit,  $P_{13}$ was calculated in \cite{Polchinski2003} in terms of Bessel functions, so that, to obtain $F_2$, we simply  replace in (\ref{eq:A}), 
$
P_{13}(z)\rightarrow P_{13}(z,Q^2)=\frac{1}{z} (Qz)^{2}(K_{0}^{2}(Qz)+K_{1}^{2}(Qz))
$.
For DVCS, states 1 and 3 are replaced by currents for an off-shell and an on-shell photon respectively,  with $P_{13}$ given by product of  unnormalized wave-functions for appropriate R-currents. We can calculate these by evaluating the R-current - graviton Witten diagram in AdS, and we get 
$
P_{13}(z) = - C\, \frac{\pi^2}{6}\,z^3\, K_1(Q z). \label{eq:p13}
$
Here $C$ is a normalization constant that can be calculated in the strict conformal limit. 
The DVCS cross section and differential cross section can then be calculated from $A(s,t)$ via 
$
\frac{d\sigma}{dt} (x,Q^2,t)= \frac{|A|^2}{16\pi s^2}  
$
and
$
\sigma (x,Q^2)=\frac{1}{16\pi s^2}  \int dt \, |A|^2 
$.

\subsection{Pomeron Kernel:}

The leading order BFKL Pomeron has  remarkable properties. It  enters into the first term in the large $N_c$ expansion with zero beta function.  Thus it is in effect the 
weak coupling cylinder graph for the  Pomeron for a large $N_c$  conformal theory, the same approximations used in the
AdS/CFT approach albeit at strong coupling. Remarkable BFKL integrability properties allows one to treat the BFKL  kernel 
as the solution to  an  $SL(2,\mathcal{C})$ conformal spin chain. Going to strong coupling, the  two gluon  exchange  evolves into a closed string
of infinitely many tightly bound  gluons, but the same underlying symmetry persists--- referred to as  M\"obius invariance  in string theory or the
isometries of the transverse $AdS_3$ impact parameter geometry.  The position of the $j$-plane cut moves
from  $j_0 = 1+ \log (2) \lambda /\pi^2$  up to $j_0 = 2- 2/\sqrt{\lambda} $.

The BPST Pomeron kernel in the $J$-plane, $G_j(t,z,z')$,
obeys a Schr\"odinger equation on $AdS_3$ space, with $j$ serving as eigenvalue for the Lorentz boost operators $M_{+-}$. 
\ignore
{$
\left[ (-\partial_u^2 - te^{-2u})/2+\sqrt{\lambda}(j-j_0) \right]G_j(t,z,z')=\delta(u-u'),
$
with $z=e^{-u}.$ }  
In the conformal limit,   
$
G_j(t,z,z')= \int_0^\infty \frac{dq^2}{2} \; \frac{J_{\tilde \Delta(j)}(zq) J_{\tilde\Delta(j)}(qz')}{q^2-t},
$
with  $\tilde\Delta(j)^2= 2\lambda (j-j_0)$. The full Pomeron kernel can then be  obtained via an inverse Mellin transform.  In the mixed-representation, one has 
\be
K(s,b,z,z')\sim -\int \frac{dj}{2\pi i} \, {\widetilde s}^j \frac{e^{-i\pi j} + 1}{\sin\pi j} \frac{e^{(2-\Delta(j))\eta}}{\sinh \eta}
\label{eq:kernel}
\ee
where $\cosh \eta$ is the chordal distance in $AdS_3$. By integrating over $\vec b$, one obtains  
 for the imaginary part of the Pomeron kernel at $t=0$
\begin{equation}
{\rm Im}\; \mathcal{K}(s, t=0, z,z') \sim \frac{s^{\textstyle j_0}}{\sqrt{\pi\mathcal{D}\log s}}\; e^{\textstyle -(\log z-\log z')^2/\mathcal{D}\log s},\label{eq:strongkernel}
\end{equation}
which exhibits diffusion in the ``size" parameter, $\log z$,  for the exchanged closed string.  This is analogous to the BFKL kernel  at weak coupling where diffusion takes place in  $\log(k_\perp)$,  the virtuality of the off shell gluon dipole. 
The diffusion constant becomes $\mathcal{D} = 2/\sqrt{g^2N_c}$ at strong coupling compared to $\mathcal{D}  = 7 \zeta(3) g^2 N_c/2 \pi^2 $ in weak coupling.
The close analogy between the weak and strong coupling Pomeron
suggests the development of a hybrid phenomenology leveraging plausible interpolations between the two extremes.

\subsection{Fit to HERA Data}

\begin{figure}
\begin{center}
\includegraphics[height=0.30 \textwidth,width=.40\textwidth]{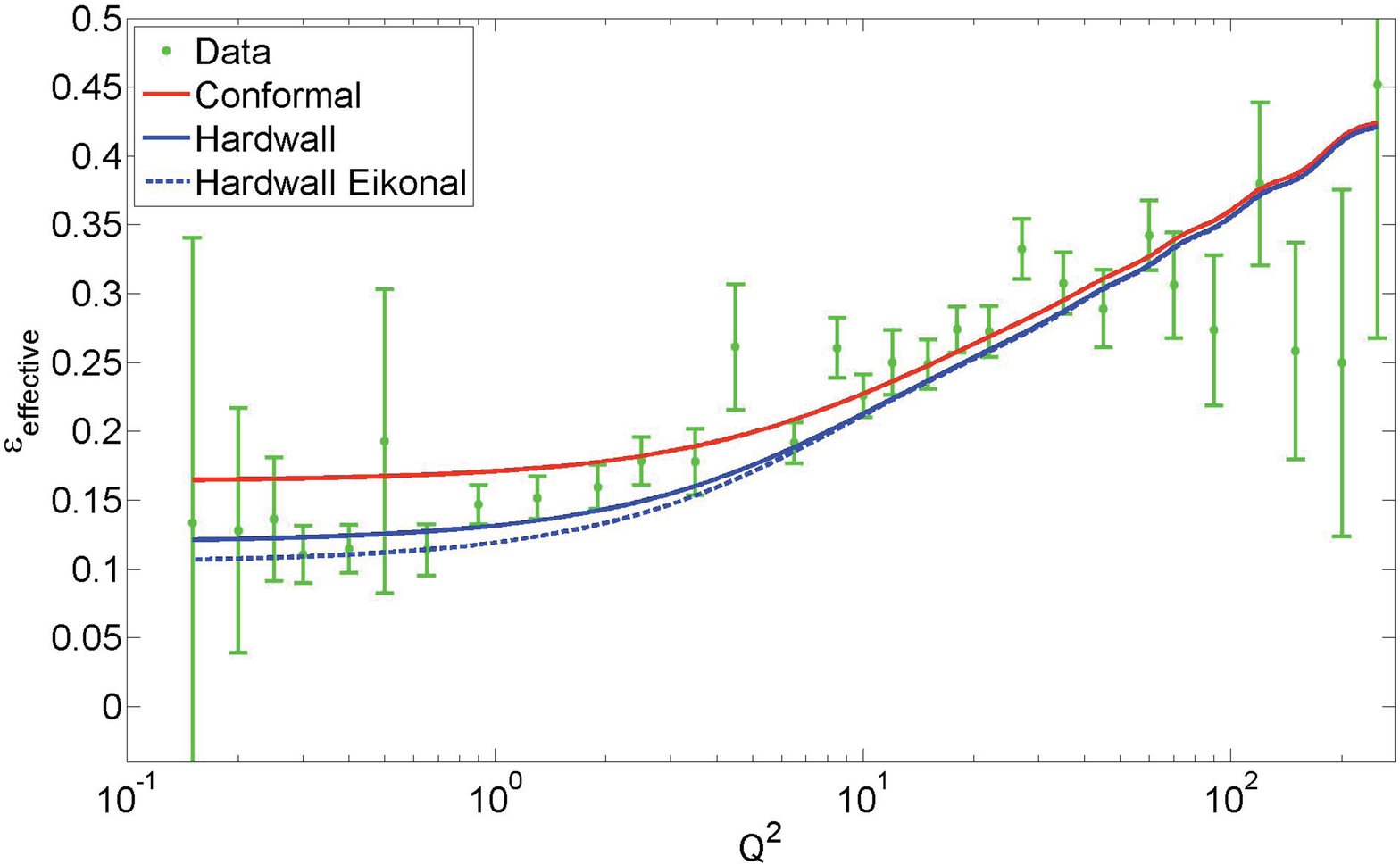}
\hskip 40pt
\includegraphics[height=0.30 \textwidth,width=.45\textwidth]{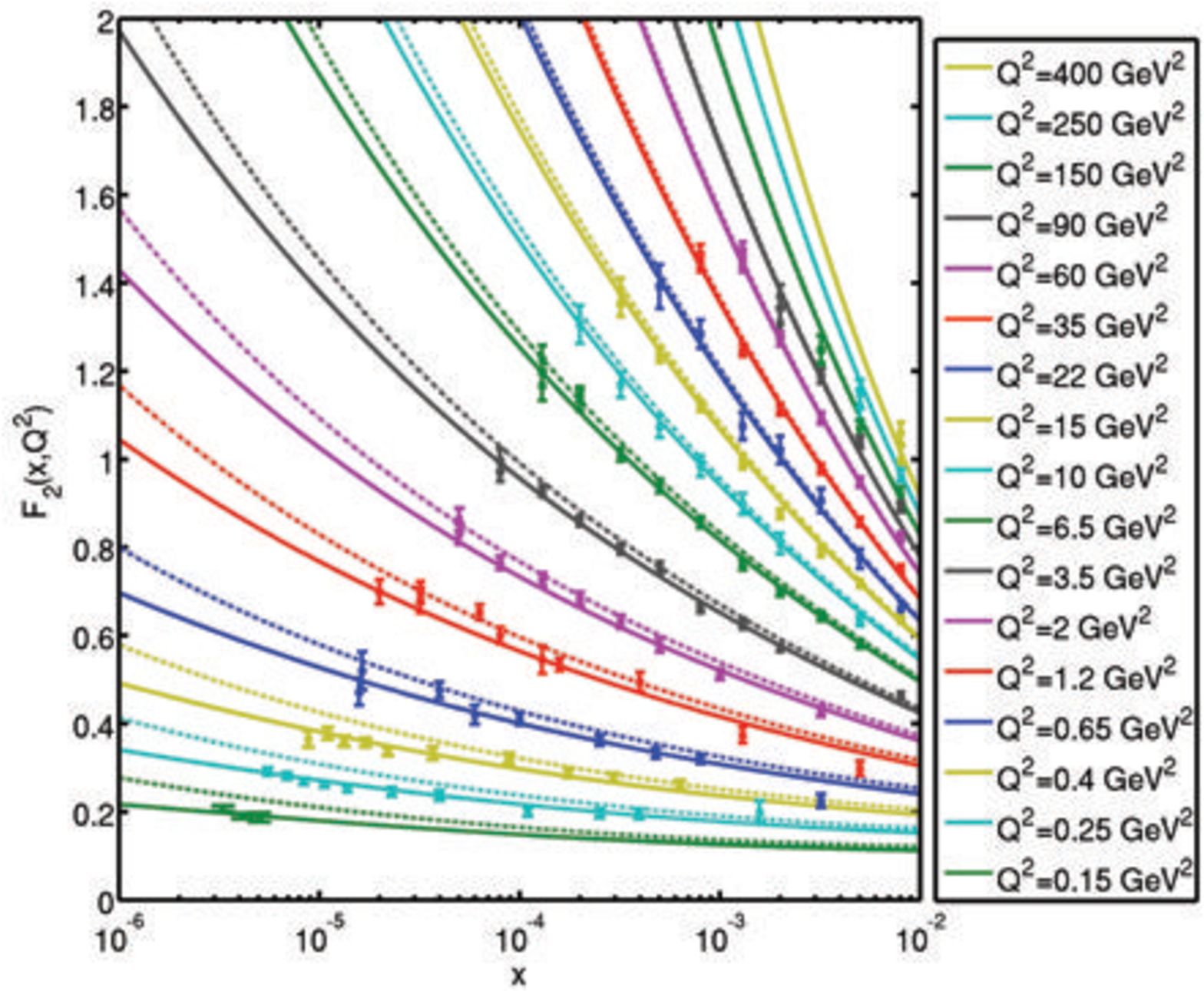}
\end{center}
\caption{On the left, (a), $Q^2$ dependence for ``effective intercept". On the right, (b), fits  by a single hardwall Pomeron vs hardwall eikonal.}
\label{fig:HERALHC}
\end{figure}

Both of these integrals, $z$ and $z'$ in (\ref{eq:A}),   remain sharply peaked, the first around $z\sim 1/Q$ and the second around the inverse proton mass, $z'\equiv 1/Q'\sim 1/m_p$. We  approximate both of them by delta functions.   Under such an ``ultra-local" approximation, all structure functions take on very simple form,  e.g, 
$$
F_{2}(x,Q^2) =\frac{g_{0}^2 }{8 \pi^{2}\lambda} \frac{Q}{Q'}\frac{e^{\textstyle (j_0 -1)\;\tau} }{\sqrt{\pi {\cal D} \tau }} \;e^{\textstyle -  (\log Q-\log Q')^2/ {\cal D} \tau}    + {\rm Confining \;\; Images}, \label{eq:f2conformalb}
$$
with  diffusion time given more precisely as $\tau =  \log ( s/QQ'\sqrt{\lambda}) =  \log (1/x) - \log (\sqrt{\lambda}  Q'/Q)$.  Here the first term is conformal and, for the hardwall, the confining effect can be expressed in terms of image charges \cite{Brower2010}. It is important to note  that taking the $s\rightarrow \infty$ limit, the single-Pomeron amplitude   grows asymptotically as $(1/x)^{j_0} \sim s^{j_0}$, thus violating the Froissart unitarity bound at very high energies. The eikonal approximation in $AdS$ space \cite{Brower2009a,Cornalba2008} plays
the role of   implementing ``saturation"  to restore  unitary via multi-Pomeron shadowing.

We have shown various comparisons of our results~\cite{Brower:2010wf} to the small-x DIS data from the combined H1 and ZEUS experiments at HERA~\cite{Aaron2010a} in Fig.~\ref{fig:HERALHC}. Both the conformal, the hard-wall model as well as the eikonalized hard-wall model can fit the data reasonably well. This can best be seen in Fig.~\ref{fig:HERALHC}a to the left  which exhibits the $Q^2$ dependence of an effective Pomeron intercept. This can be understood  as a consequence of diffusion. However, it is important to observe that  the hard-wall model provides a much better fit than the conformal result  for $Q^2$ less than $2\sim 3 $ $GeV^2$. 
The best fit to data is obtained using the hard-wall eikonal model, with a $\chi^2 = 1.04$.  This is clearly shown by Fig.~\ref{fig:HERALHC}b to the right, where we present a comparison of the relative importance of confinement versus eikonal at the current energies.
We observe  that the   transition scale $Q_{c}^2(x)$  from conformal to confinement increases with  $1/x$, and it comes before saturation effect becomes important.  For more details, see Ref. ~\cite{Brower:2010wf}.
\begin{figure}
\begin{center}
\includegraphics[height=0.25 \textwidth,width=.3\textwidth]{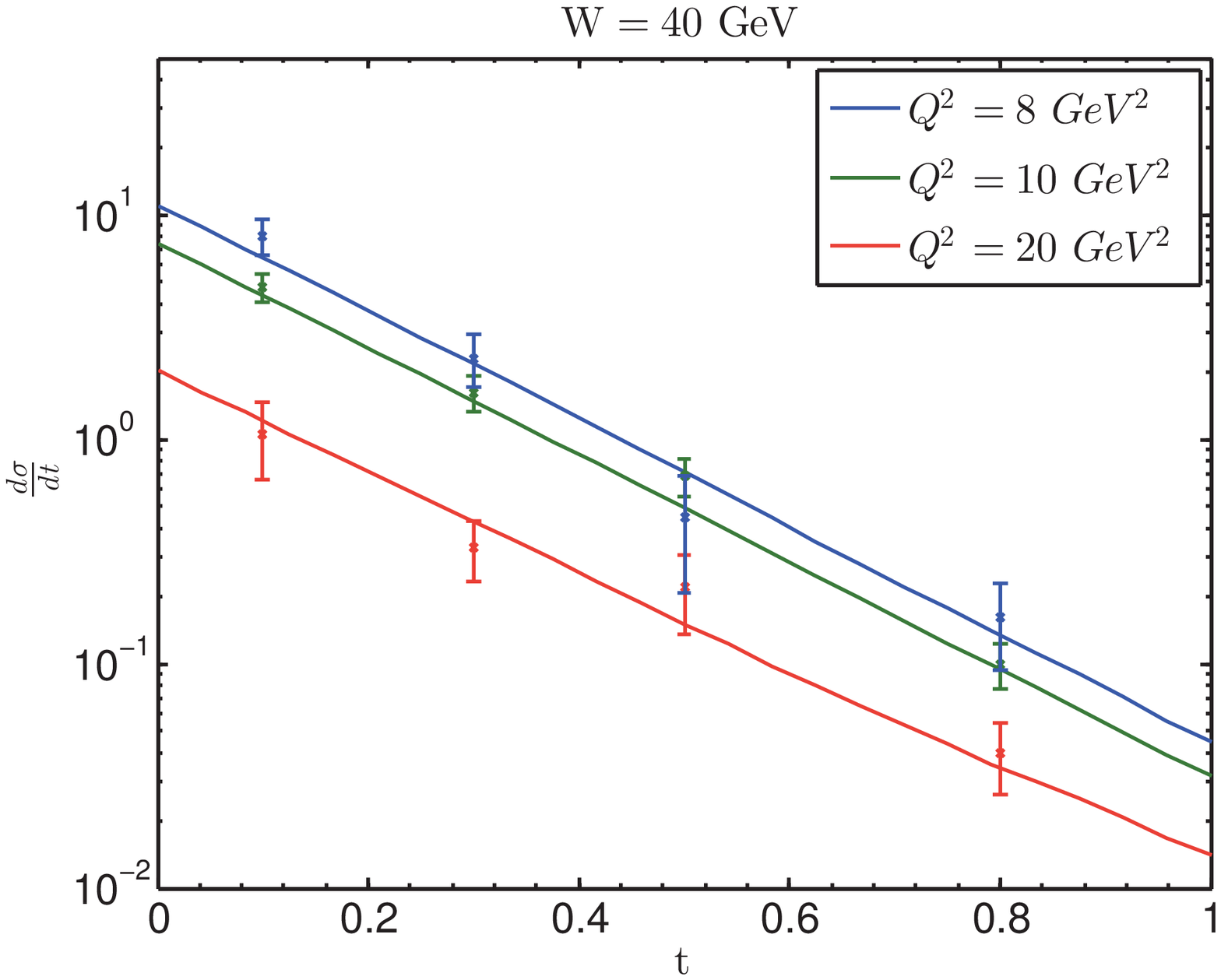}
\includegraphics[height=0.25 \textwidth,width=.3\textwidth]{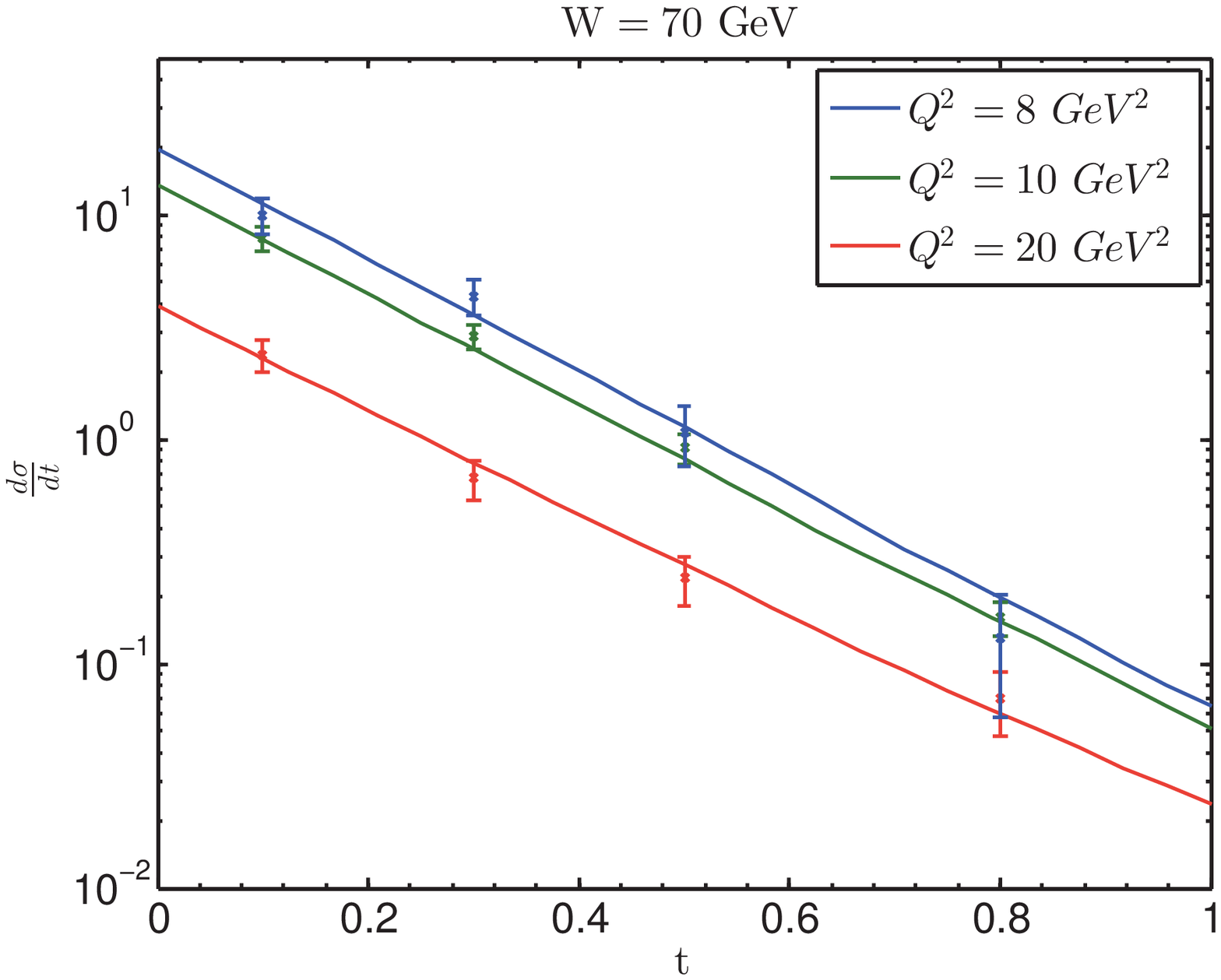}
\includegraphics[height=0.25 \textwidth,width=.3\textwidth]{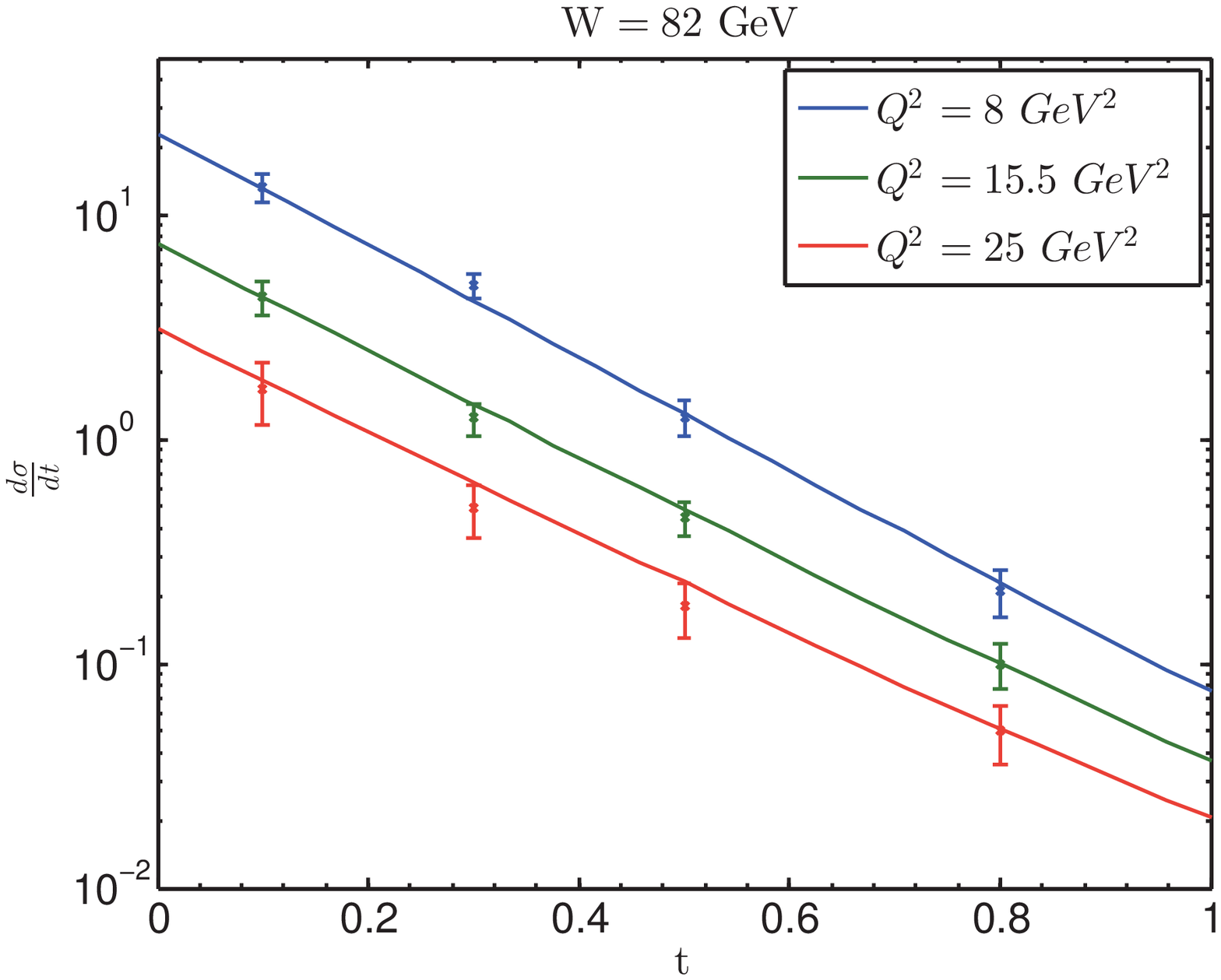}
\hskip 0.5cm 
\includegraphics[height=0.25 \textwidth,width=.3\textwidth]{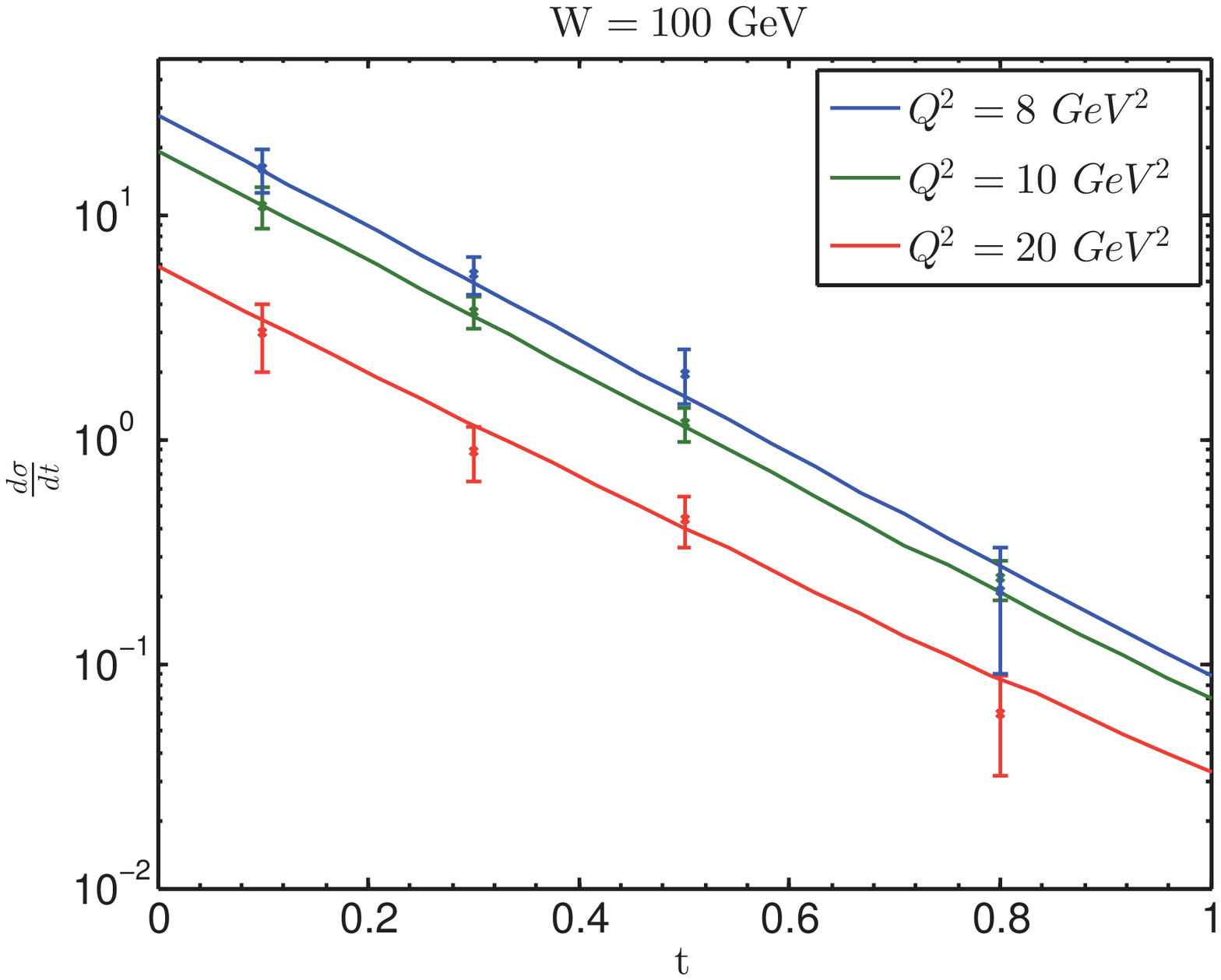}
\includegraphics[height=0.25 \textwidth,width=.3\textwidth]{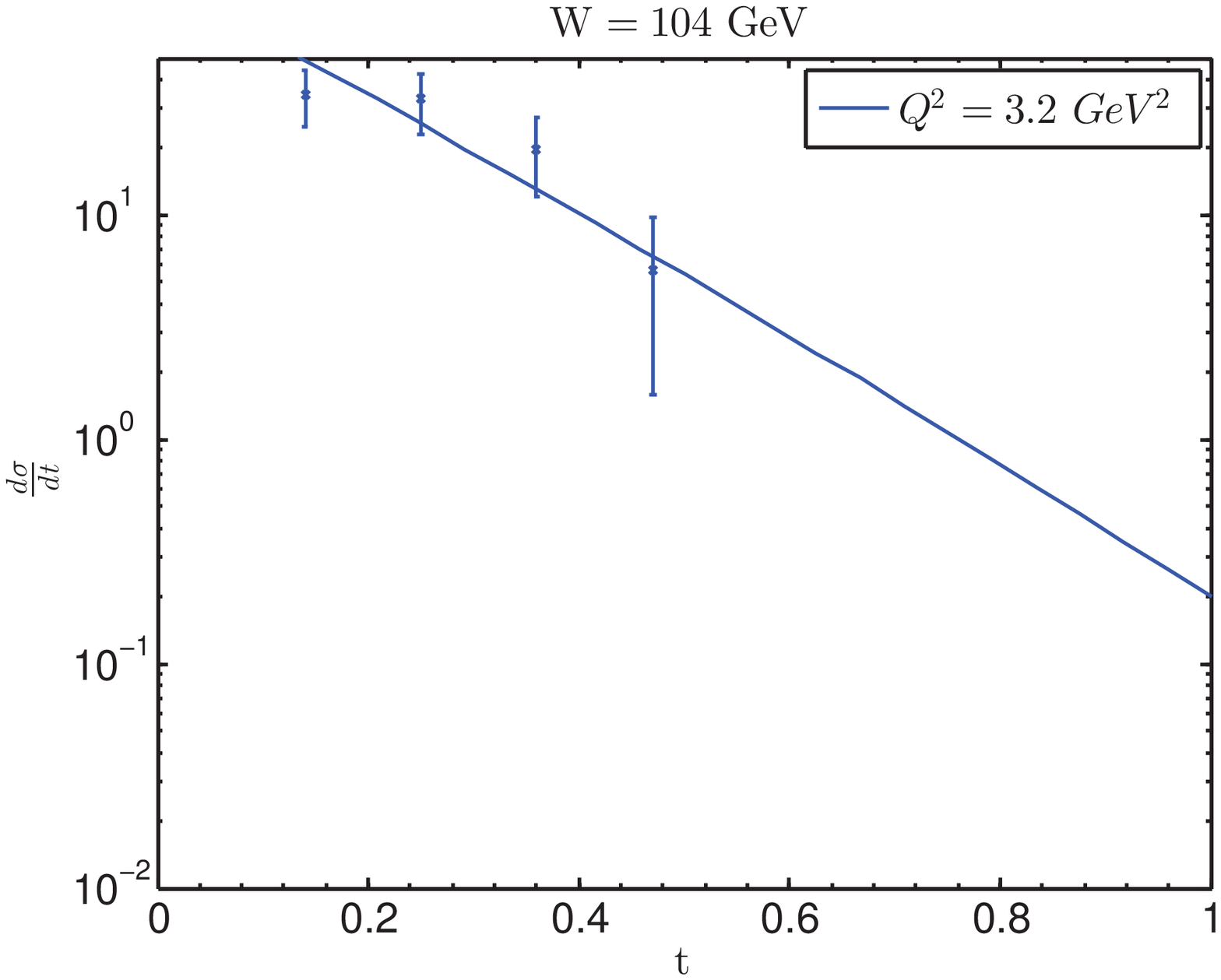}
\includegraphics[height=0.25 \textwidth,width=.3\textwidth]{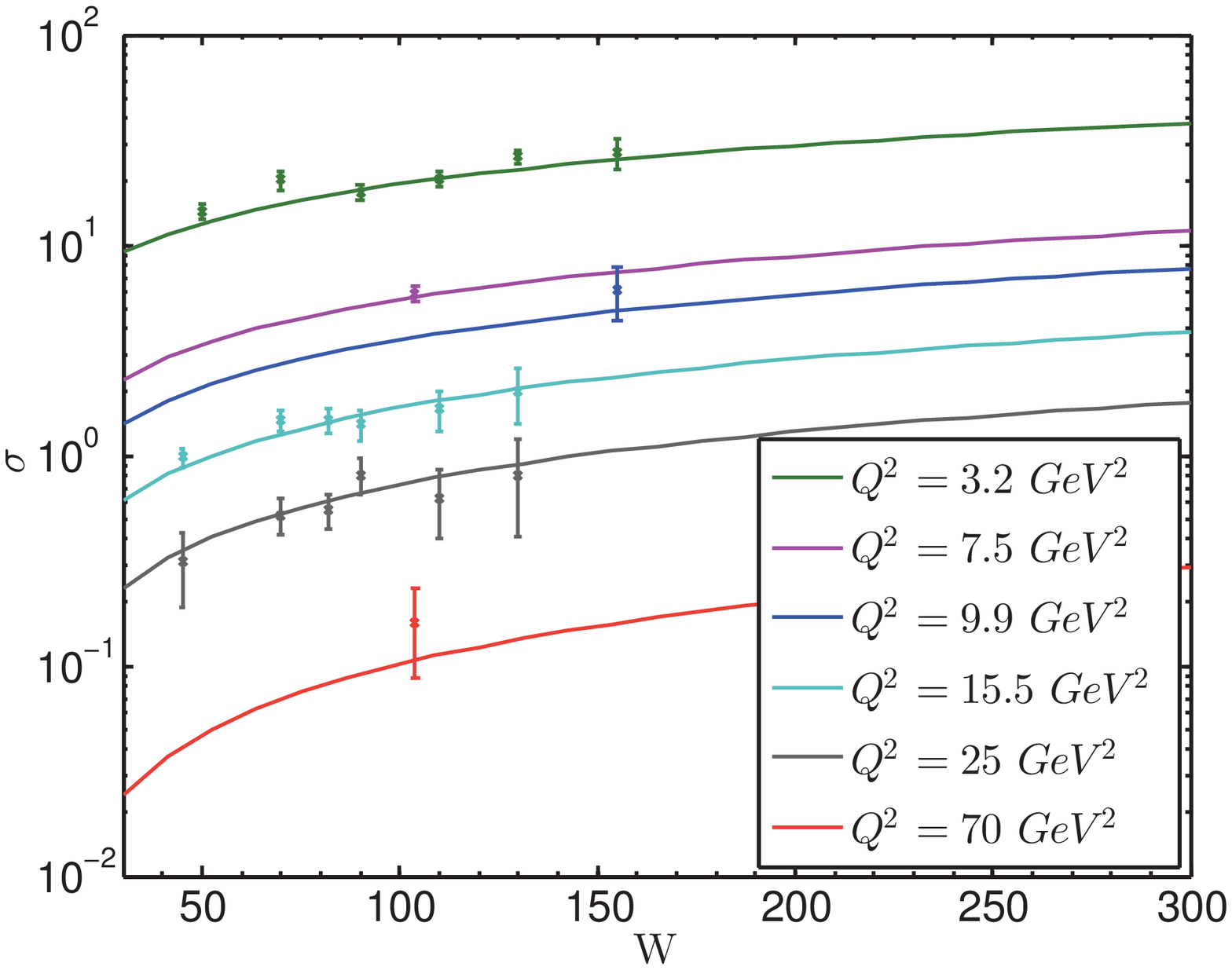}
\end{center}
\caption{Fits by the hard wall Pomeron to HERA data. The first 5 correspond to the differential cross section, and the last one to the cross section.}
\label{fig:dvcs_fit}
\end{figure}
\ignore{
\begin{figure}
\begin{center}
\includegraphics[height=0.125 \textwidth,width=.15\textwidth]{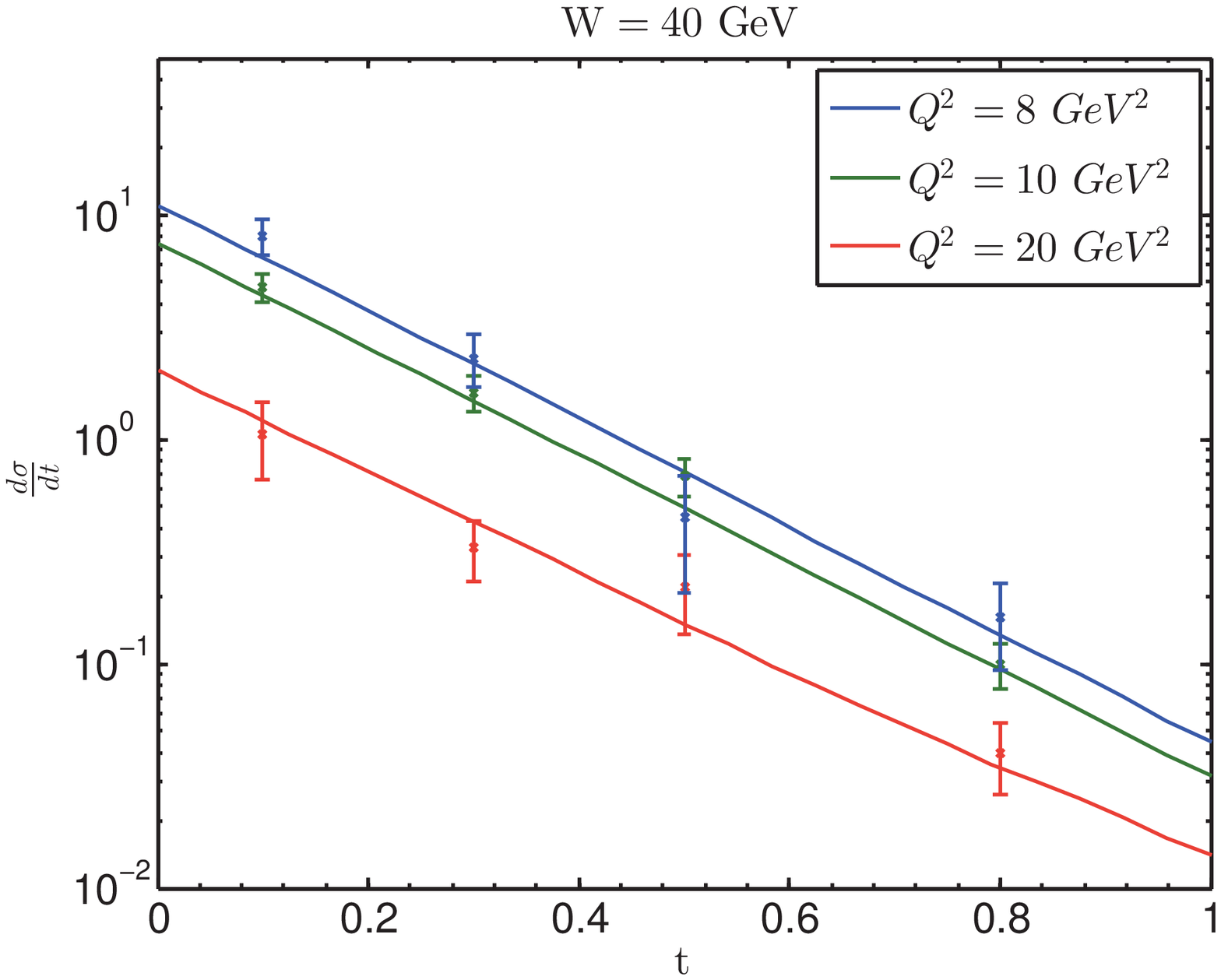}
\includegraphics[height=0.125 \textwidth,width=.15\textwidth]{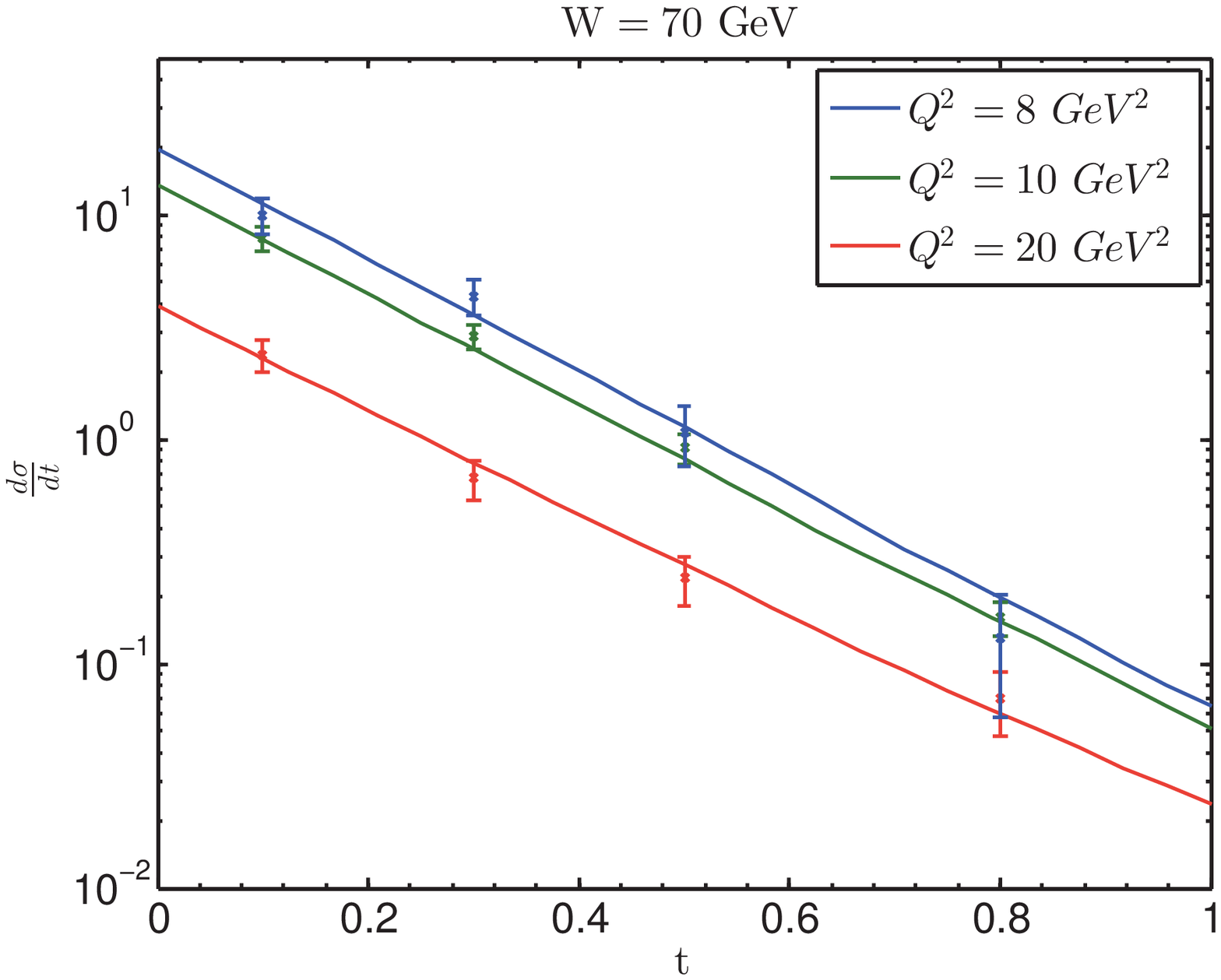}
\includegraphics[height=0.125 \textwidth,width=.15\textwidth]{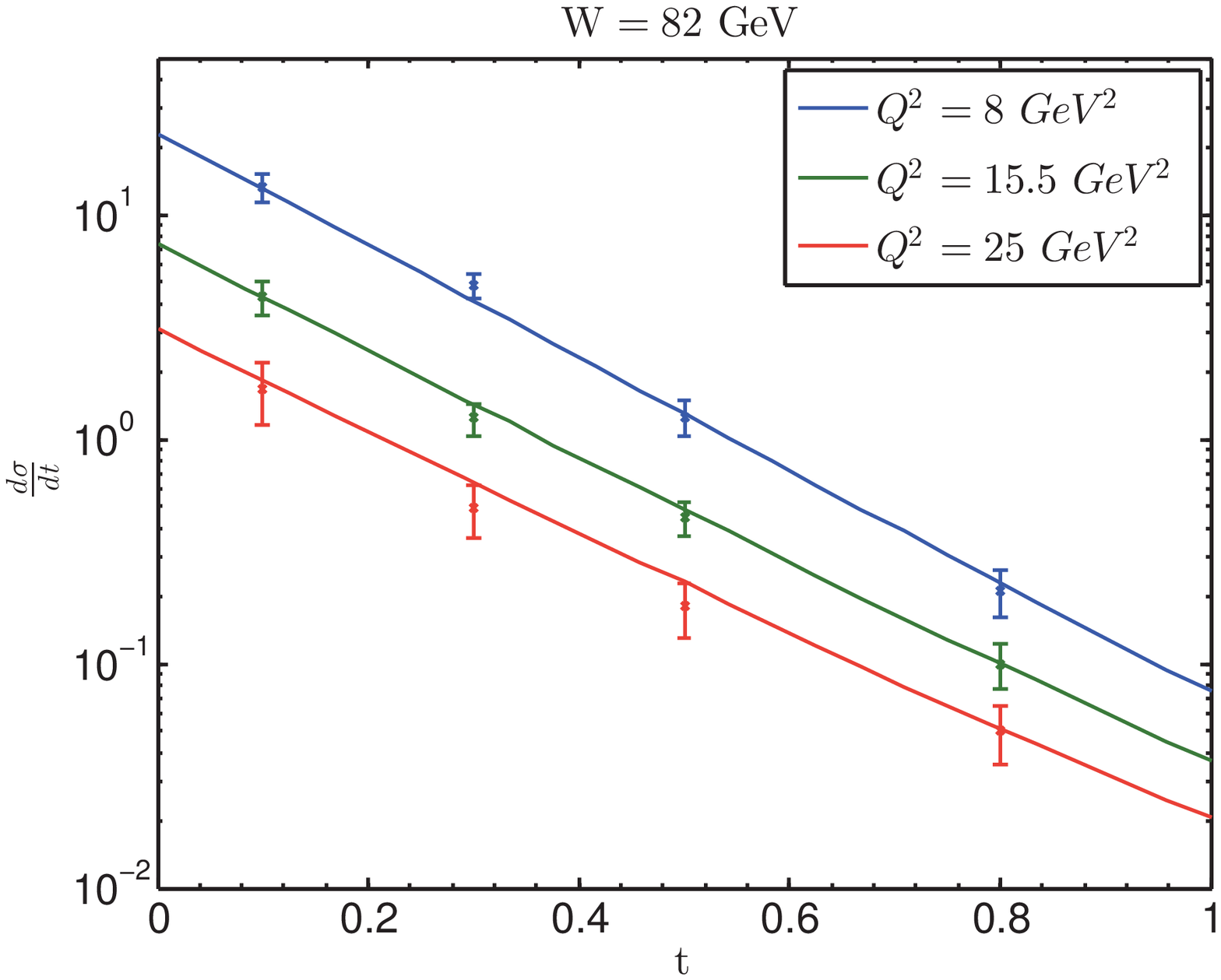}
\includegraphics[height=0.125 \textwidth,width=.15\textwidth]{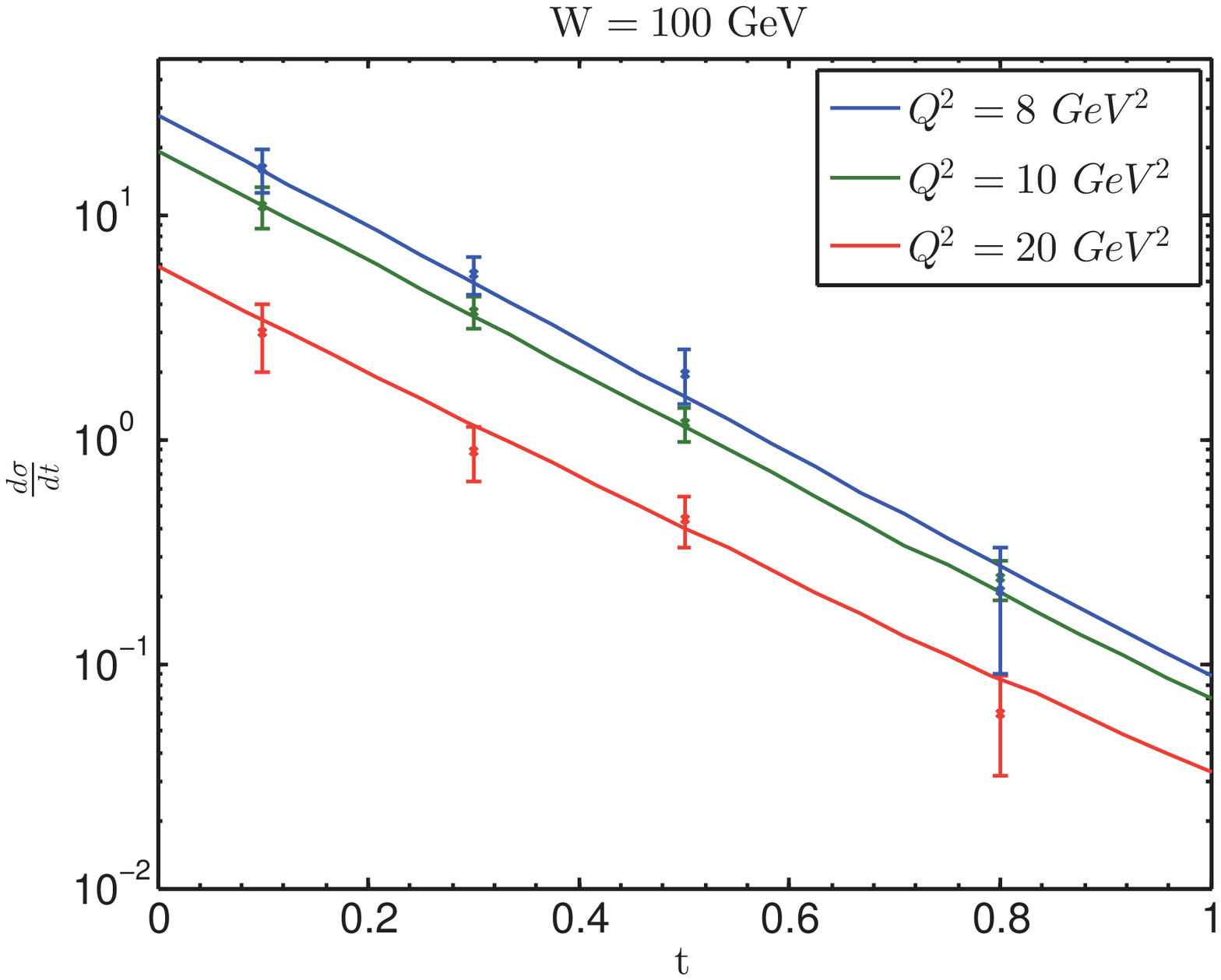}
\includegraphics[height=0.125 \textwidth,width=.15\textwidth]{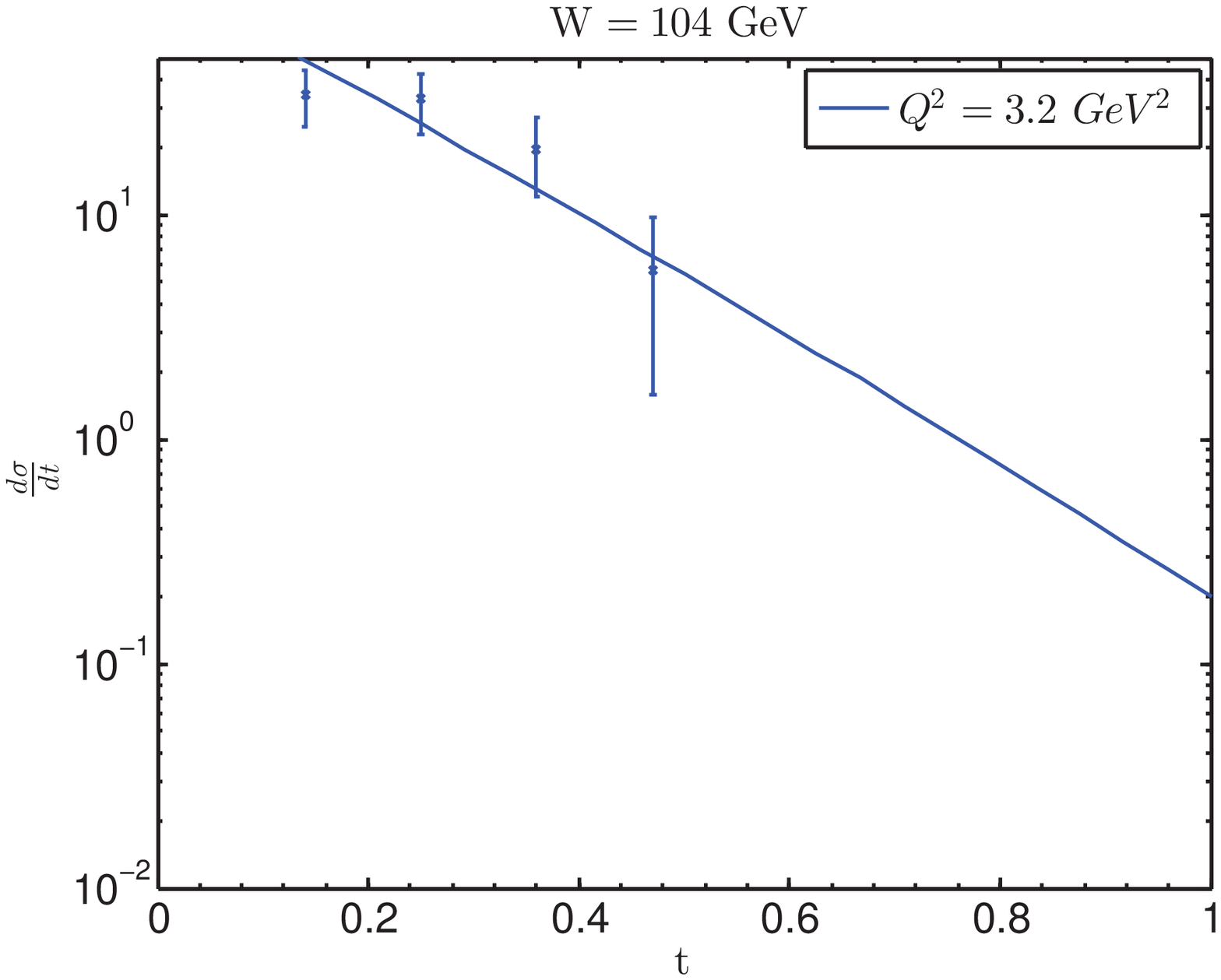}
\includegraphics[height=0.125 \textwidth,width=.15\textwidth]{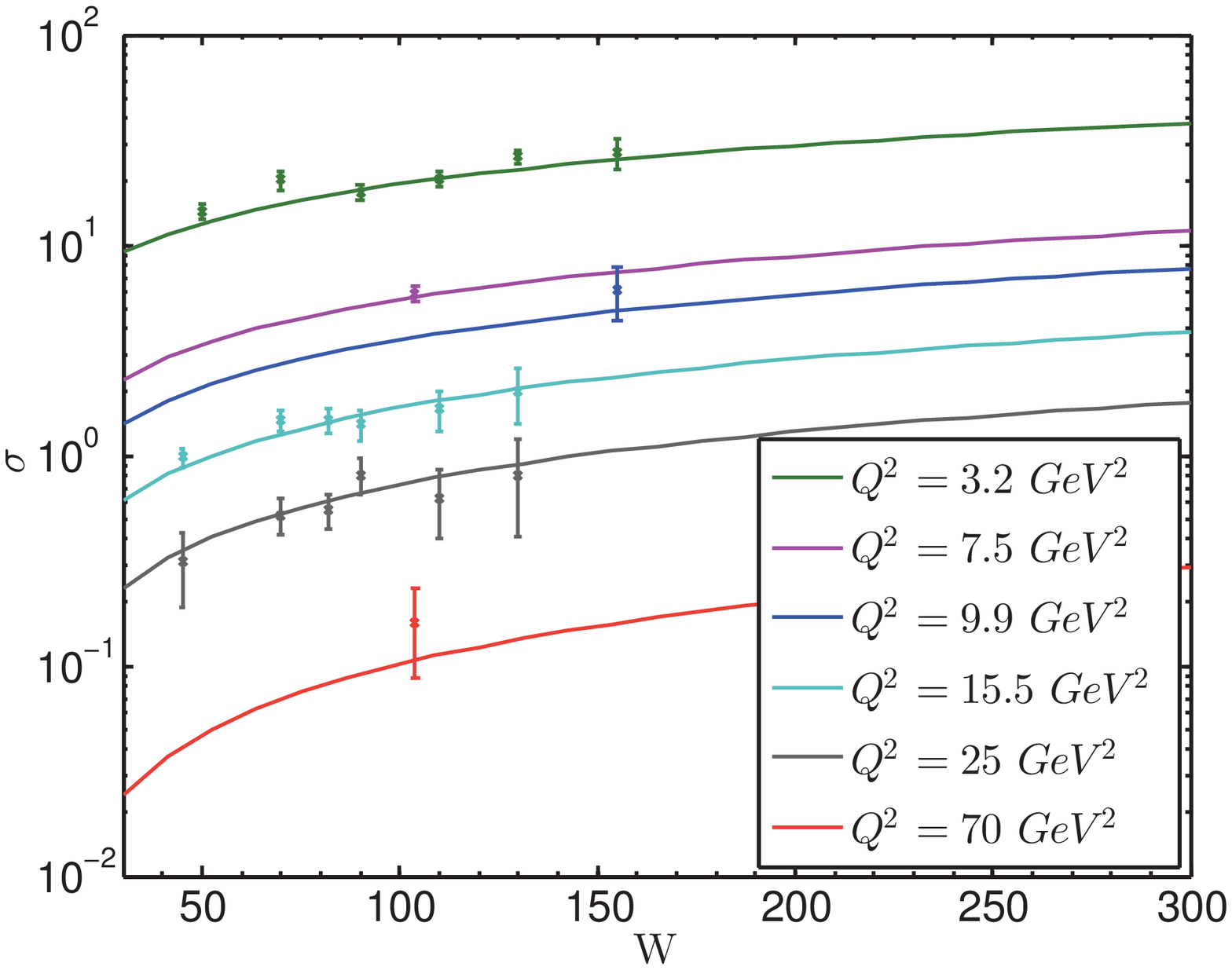}
\end{center}
\caption{Fits by the hard wall Pomeron to HERA data. The first 5 correspond to the differential cross section, and the last one to the cross section.}
\label{fig:dvcs_fit}
\end{figure}
}

We now compare our model to the measurements at HERA. Related papers using AdS/CFT correspondence applied to DVCS can be found in \cite{Costa:2012fw}. We use conformal form  for the photon wave functions and a delta function for the proton. Note that Eq. (\ref{eq:strongkernel}) is for the conformal model, and the hard wall expression would include another term with the contribution due to the presence of the hard wall.  We obtain a good agreement with experiment, with $\chi^2$ varying from $0.51 - 1.33$ depending on the particular data and model we are considering. We find that confinement starts to play a role at small $|t|$, and the hardwall fits the data better in this region. \ignore{Explicitly, the parameter values we get for the hard wall model are $g_0^2 = 2.46\, \pm  0.70\,, z_* = 3.35 \pm 0.41\ {\rm GeV}^{-1},\ \rho = 0.712 \pm 0.038\,,\ z_0 = 4.44 \pm 0.82\ {\rm GeV}^{-1}$ for the differential cross section, and $g_0^2 = 6.65 \pm 2.30\,, z_* = 4.86 \pm 2.87\ {\rm GeV}^{-1} ,\ \rho = 0.811 \pm 0.036\,,\  z_0 = 8.14 \pm 2.96\ {\rm GeV}^{-1},$ with $\chi^2_{d.o.f}=0.51$ and $1.03$ respectively.} In Fig.~\ref{fig:dvcs_fit} we present the plots for our fits.    (See \cite{Costa:2012fw} for details.)

\section{Regge Limit in CFT, Anomalous Dimensions, Conformal Pomeron and Odderon}
\label{sec:CFT}
Let us return to take a closer  look at the Regge limit for a CFT. Consider a connected 4-point correlation function for a scalar field $\phi(x)$ of dimension $\Delta_0$, ${\cal A}=\langle \phi(x_1) \phi(x_2)  \phi(x_3)  \phi(x_4)\rangle$, which can be expressed as
$
{\cal A} = x_{12}^{-\Delta_0} x_{34}^{-\Delta_0} F(u,v)
$. 
Here $x^2_{ij}=(x_1-x_j)^2$, and  ($u$, $v$) are two conformal invariant cross ratios, $u=x_{12}^2x_{34}^2/x_{14}^2x_{23}^2$, and $v=x_{13}^2x_{24}^2/x_{14}^2x_{34}^2$. It can formally be expanded in an operator product expansion, which, in turn, can be expressed in a conformal partial-wave expansion, 
$
F(u,v)=\sum_k \sum_{j ,\Delta_{k,j}} C_{k,j} G_j(u,v; \Delta_{k,j})
$.
Here $G_j(u,v; \Delta_{k,j})$ are the ``conformal blocks", associated with a primary,  ${\cal O}_{k,j}$, which includes contributions from its descendants. For each $j$, there can be many such primaries, labelled by the index $k$.  Of these, we will be interested in the family of  primaries, ${\cal O}_{G,j}$, for even $j$,  which interpolate with the energy-momentum tensor at $j=2$. Since we are dealing with a CFT with AdS dual, this family of primaries are that dual to string modes associated with the graviton, with spin $j=2,4,\cdots$, i.e., those given by Eq. (\ref{eq:GAD}). The challenge is to find the associated anomalous dimensions in the strong coupling limit.

In a coordinate treatment, it can be shown that the Regge limit corresponds to the simultaneous limits of $u\rightarrow 0$ and  $v\rightarrow 1$, with $\xi=(1-v)/\sqrt u$ fixed. In the Euclidean region, $G_j(u,v; \Delta_{k,j})\sim u^{ \Delta_{k,j}/2}\sim 0$.       However, the Regge limit of interest is defined in the Minkowski region, where 
$$
G_j(u,v; \Delta_{k,j})\sim u^{(1-j)/2} H(\xi; j,  \Delta_{k,j}) 
$$ 
diverges as $u\rightarrow 0$  The $j$-sum can be performed via a Sommerfeld-Watson resummation. After taking into account the analytic structure, e.g.,  Eq. (\ref{eq:j-Delta}),  one finds 
$
F(u,v)\sim u^{(1-j_0)/2}H(\xi; j_0,  \Delta_{k,j_0})
$.
With $\sqrt u\sim 1/s$ and  $\xi$ identified with the chordal distance, this  precisely corresponds to that obtained earlier for the conformal Pomeron kernel given in a mixed representation, e.g., Eqs. ({\ref{eq:kernel}) and (\ref{eq:strongkernel})~\cite{Brower2007,Brower2009,Cornalba:2007fs,Costa:2012cb}.

Let us now return to the determination of the Pomeron intercept $j_0$; as discussed earlier, $j_0$ is fixed by knowing $\Delta_{G,j}$, analytically continued to small $j$.
 In a remarkable paper~\cite{Basso:2011rs}, Basso, taking advantage of many recent ingenious calculations of anomalous dimensions for ${\cal N}=4$ YM based on integratibility,  has been able to generalize the work of \cite{Brower2007} so that the coefficient  $\alpha_1(\lambda)$ in Eq.~(\ref{eq:BPST-Basso}), is known exactly, for all $\lambda$. Furthermore, coefficient for next few orders in the $(j-2)$-expansion  are also known through these studies. Taking advantage of the supersymmetry, it is sufficient to work with the scalars and investigate anomalous dimensions of the corresponding Konishi multiplets. These can be symbolically represented by $tr  D^S Z^J$ + mixing, where $D$ is a covariant derivative and $Z$ is a complex scalar.  (In \cite{Basso:2011rs}, $J$ is the twist. We shall use $\tau$ instead.) For these operators, conformal dimensions admit a small $S$ expansion,
$
\Delta = \tau + \alpha_1 (\lambda,\tau) S + O(S^2)
$.

In making connection with our problem at hand, we need to identify $\Delta$ above with $\Delta-2$, $\tau\rightarrow 2$, and $S\rightarrow j-2$, leading to Eq. (\ref{eq:BPST-Basso}).
Maintaining the symmetry in $\Delta\leftrightarrow 4-\Delta$, and keeping enough terms to evaluate up to $O(\lambda^{5/2})$, we have
\begin{equation}\label{eq:Basso}
	\begin{aligned}	
(\Delta-2)^2 =   \tau^2  &+  \Big( 2\sqrt\lambda  -1 +\frac{\tau^2-\frac{1}{4}}{\sqrt \lambda}+\cdots \Big ) S  
+ \Big( \frac{3}{2} - \frac{b_0}{\sqrt \lambda} +\frac{b_1}{\lambda}+\cdots \Big ) S^2\\
 &+(- \frac{3}{8\sqrt\lambda}+\cdots  )S^3 + O(S^4)\nn
	\end{aligned}
\end{equation}
where the $b_i$ can be fixed by calculating the energy of spinning string in $AdS_5\times S^5$ in a loop expansion. After inverting, and evaluating the minimum of $j(\Delta)$ at $\Delta=2$, one obtains~\cite{Costa:2012cb,Kotikov:2013xu} (note: the $b_0$ term is completely determined by the the one-loop calculation)
$$
\alpha_P=j_{0}=2 -\frac{2 }{\lambda^{1/2}}-\frac{1}{\lambda} +\frac{1}{ 4\lambda^{3/2}}+\frac{6\zeta(3)+2 }{\lambda^2}+\frac{-2b_1+9\zeta(3)+\frac{245}{64}}{\lambda^5/2}+\cdots.
\label{eq:P-intercept}
 $$
 
 A similar analysis can also be applied for Odderons.   Recall that, from AdS/CFT, it has been shown that Odderons can be identified with modes associated with the anti-symmetric Kalb-Ramond fields in AdS. We note, in particular,  the spin approaches $j=1$ in the super-gravity limit. 
It follows that  a similar expansion in $\lambda^{-1/2}$ can also be carried out, leading to an expansion where 
 $
\alpha_O= j^{-}_0= 1 +\frac{a^-_{1}}{\lambda^{1/2}}+\frac{a^-_2}{\lambda} +\frac{a^-_3}{ \lambda^{3/2}}+\frac{a^-_4}{\lambda^2}+\cdots.
$.
In \cite{Brower2009}, two Odderon solutions were found. Solution-A has $a^{-}_{1,A}= -8$.  The second solution, solution-B, surprisingly remains at 1 in the large $\lambda$ limit, i.e., $a^{-}_{1,B}=0$. For solution-B, one  immediate question is whether this pattern will survive at higher order.

  We have recently  extended the analysis of \cite{Brower2009} to higher order, leading to  a similar expansion as Eq. (\ref{eq:P-intercept}). 
  For solution-A,  we find
   $$
\alpha_{O,A}= j^{-}_{0,A}= 1 -\frac{8}{\lambda^{1/2}}-\frac{4}{\lambda} +\frac{13}{ \lambda^{3/2}}+\frac{96\zeta(3)+41}{\lambda^2}+\frac{\frac{-81}{256}(45+32b_1-144\zeta(3))}{\lambda^{5/2}}+\cdots.
$$
 \ignore{ It should be noted that, as $\lambda$ decreases, $\alpha_{O,A}$ should approach 1. It appears that, from strong coupling, this approach will be highly oscillatory.  }
 Let us turn next to solution $B$.   In order to match the vanishing first order correction, we find that $a_{i,B}^{-} =0$ for all $i=1,2,\cdots$ recursively. This leads to a surprising conclusion that 
$$
\alpha_O=1
$$
to all orders in $1/\sqrt\lambda$ in a strong coupling treatment. It is also interesting that this finding is consistent with that reported from a weak-coupling analysis~\footnote{See \cite{Brower2009} for references cited.}.  More detailed analysis will be presented  in a forthcoming report~\footnote{Richard Brower, Miguel Costa, Marko Djuric, Timothy Raben and  Chung-I Tan, ``Conformal Pomeron and Odderon", to appear.}.

\bibliographystyle{unsrt}
\bibliography{lowx}

\begin{thebibliography}{10}

\bibitem{Brower2007}
Richard~C. Brower, Joseph Polchinski, Matthew~J. Strassler, and Chung-I Tan.
\newblock {The Pomeron and gauge/string duality}.
\newblock {\em JHEP}, 0712:005, 2007.

\bibitem{Brower2009a}
Richard~C. Brower, Matthew~J. Strassler, and Chung-I Tan.
\newblock {On The Pomeron at Large 't Hooft Coupling}.
\newblock {\em JHEP}, 0903:092, 2009.

\bibitem{Brower2009}
Richard~C. Brower, Marko Djuric, and Chung-I Tan.
\newblock {Odderon in gauge/string duality}.
\newblock {\em JHEP}, 0907:063, 2009.

\bibitem{Aaron2010a}
F.D. Aaron et~al.
\newblock {Combined Measurement and QCD Analysis of the Inclusive e+- p
  Scattering Cross Sections at HERA}.
\newblock {\em JHEP}, 1001:109, 2010.

\bibitem{Brower2010}
Richard~C. Brower, Marko Djuric, Ina Sarcevic, and Chung-I Tan.
\newblock {String-Gauge Dual Description of Deep Inelastic Scattering at
  Small-$x$}.
\newblock {\em JHEP}, 1011:051, 2010.

\bibitem{Costa:2012fw}
Miguel~S. Costa and Marko Djuric.
\newblock {Deeply Virtual Compton Scattering from Gauge/Gravity Duality}.
\newblock {\em Phys.Rev.}, D86:016009, 2012.

\bibitem{Brower2012}
Richard~C. Brower, Marko Djuric, and Chung-I Tan.
\newblock {Diffractive Higgs Production by AdS Pomeron Fusion}.
\newblock {\em JHEP}, 1209:097, 2012.

\bibitem{Kotikov:2004er}
A.V. Kotikov, L.N. Lipatov, A.I. Onishchenko, and V.N. Velizhanin.
\newblock {Three loop universal anomalous dimension of the Wilson operators in
  N=4 SUSY Yang-Mills model}.
\newblock {\em Phys.Lett.}, B595:521--529, 2004.

\bibitem{Polchinski2003}
Joseph Polchinski and Matthew~J. Strassler.
\newblock {Deep inelastic scattering and gauge / string duality}.
\newblock {\em JHEP}, 0305:012, 2003.

\bibitem{Cornalba2008}
Lorenzo Cornalba, Miguel~S. Costa, and Joao Penedones.
\newblock {Eikonal Methods in AdS/CFT: BFKL Pomeron at Weak Coupling}.
\newblock {\em JHEP}, 0806:048, 2008.

\bibitem{Brower:2010wf}
Richard~C. Brower, Marko Djuric, Ina Sarcevic, and Chung-I Tan.
\newblock {String-Gauge Dual Description of Deep Inelastic Scattering at
  Small-$x$}.
\newblock {\em JHEP}, 1011:051, 2010.

\bibitem{Cornalba:2007fs}
Lorenzo Cornalba.
\newblock {Eikonal methods in AdS/CFT: Regge theory and multi-reggeon
  exchange}.
\newblock 2007.

\bibitem{Costa:2012cb}
Miguel~S. Costa, Vasco Goncalves, and Joao Penedones.
\newblock {Conformal Regge theory}.
\newblock {\em JHEP}, 1212:091, 2012.

\bibitem{Basso:2011rs}
B.~Basso.
\newblock {An exact slope for AdS/CFT}.
\newblock 2011.

\bibitem{Kotikov:2013xu}
A.V. Kotikov and L.N. Lipatov.
\newblock {Pomeron in the N=4 supersymmetric gauge model at strong couplings}.
\newblock {\em Nucl.Phys.}, B874:889--904, 2013.

\end{thebibliography}


\end{document}